\documentclass{pnastwo}

\usepackage[dvips]{graphicx}

\usepackage{amssymb,amsfonts,amsmath}

\contributor{Submitted to Proceedings
of the National Academy of Sciences of the United States of America}
\url{www.pnas.org/cgi/doi/10.1073/pnas.0709640104}
\copyrightyear{2008}
\issuedate{Issue Date}
\volume{Volume}
\issuenumber{Issue Number}

\begin{document}

\title{Supernova-driven outflows and chemical evolution of dwarf spheroidal galaxies}

\author{Yong-Zhong Qian\affil{1}{School of Physics and Astronomy, University of
Minnesota, Minneapolis, MN 55455}\and
Gerald J. Wasserburg\affil{2}{The Lunatic Asylum, Division of Geological and 
Planetary Sciences, California Institute of Technology, Pasadena, CA 91125}}

\contributor{Submitted to Proceedings of the National Academy of Sciences
of the United States of America}
\maketitle

\begin{article}

\begin{abstract} 
We present a general phenomenological model for the metallicity distribution (MD) 
in terms of [Fe/H] for dwarf spheroidal galaxies (dSphs). These galaxies appear to 
have stopped accreting gas from the intergalactic medium
and are fossilized systems with their stars undergoing slow internal evolution. 
For a wide variety of infall histories of unprocessed baryonic matter
to feed star formation, most of the observed MDs can be well described by our model. 
The key requirement is that the fraction of the gas mass lost by supernova-driven
outflows is close to unity. This model also predicts a relationship between
the total stellar mass and the mean metallicity for dSphs in accord with
properties of their dark matter halos. 
The model further predicts as a natural consequence that
the abundance ratios [E/Fe] for elements such as O, Mg, and Si decrease 
for stellar populations at the higher end of the [Fe/H] range in a dSph.
We show that for infall rates far below the net rate of gas loss to star
formation and outflows, the MD in our model
is very sharply peaked at one [Fe/H] value, 
similar to what is observed in most globular clusters. This suggests that
globular clusters may be end members of the same family as dSphs.
\end{abstract}


\dropcap{I}n this paper we show that supernova-driven gas outflows play a prominent role
in the chemical evolution of dwarf spheroidal galaxies (dSphs).
In the framework of hierarchical structure formation based on the cold dark matter
cosmology, dwarf galaxies are the building blocks of large galaxies such as the Milky Way.
In support of this picture, some recent observations showed that elemental abundances 
in dSphs of the Local Group match those in the Milky Way halo at low metallicities
(e.g., \cite{frebel10,simon10,cohen10,let,taf}; see \cite{tol} for a review of earlier works). 
It is expected that detailed studies of chemical evolution of dwarf 
galaxies can shed important light on the formation and evolution of the Milky Way in particular
and large galaxies in general. Here we present an analysis of the evolution of 
[Fe/H]~$=\log({\rm Fe/H})-\log({\rm Fe/H})_\odot$ focussing on dSphs. 
The approach is a 
phenomenological one that takes into account infall of gas into the dark matter halos
associated with these galaxies, star formation (SF) within the accumulated gas, and 
outflows driven by supernova (SN) explosions. 
The sources for production of Fe are core-collapse SNe
(CCSNe) from progenitors of 8--$100\,M_\odot$ and Type Ia SNe (SNe Ia)
associated with stars of lower masses in binaries.
Observations require that some SNe Ia must form early along with CCSNe 
without a significant delay. It will be shown that there is a direct and
simple connection between the metallicity distribution (MD) for a given dSph
and two parameters $\lambda_{\rm Fe}/\lambda$ and $\alpha$. The
ratio refers to the net rate $\lambda_{\rm Fe}$ of Fe production 
and the net rate $\lambda$ of gas loss to SF and 
SN-driven outflows, and $\alpha$ indicates the promptness for reaching peak 
infall rates. This model explicitly predicts the ratio of the stellar mass in the dSph 
to the total mass of the host dark matter halo. 

Phenomenological models for chemical evolution have a long history
(e.g., \cite{lb75}) and were applied to dSphs previously 
(e.g., \cite{lm04,lm07,lm10}). Dynamic models for dSphs including dark matter 
were also studied (e.g., \cite{marc,revaz09}). The first effort was made in \cite{silk} 
to reconcile models of hierarchical structure formation involving dark matter halos 
with the then-available luminosity-radius-metallicity relationships for dwarf
galaxies. There it was shown that SN-driven outflows could explain 
the observed trends. Recent observations \cite{kirby11a}
give results for eight dSphs with rather detailed structure of their 
MDs and provide a basis for exploring models of 
their chemical evolution (e.g., \cite{kirby11b}). 
Among the key issues that we try to address are MDs exhibited by
stellar populations of dSphs. Our general approach follows that of 
Lynden-Bell described in his incisive and excellent article on ``theories''
of the chemical evolution of galaxies \cite{lb92}.
It will be shown that the metallicity at which the MD peaks is directly related 
to the efficiency of SN-driven outflows and that the MD of a dSph 
and the relationship between the stellar mass and the mean metallicity
for these galaxies are direct consequences of the model. These results 
are in strong support of those of \cite{dekel}, where earlier and less 
precise data on metallicities of dwarf galaxies were used to address this problem. 

In our approach, we consider evolution of Fe in a homogeneous system 
of condensed gas governed by
\begin{eqnarray}
\frac{dM_g}{dt}&=&\left(\frac{dM_g}{dt}\right)_{\rm in}-\psi(t)-F_{\rm out}(t),
\label{eq-dmgdt}\\
\frac{dM_{\rm Fe}}{dt}&=&P_{\rm Fe}(t)-\frac{M_{\rm Fe}(t)}{M_g(t)}[\psi(t)+F_{\rm out}(t)],
\label{eq-dmfedt}
\end{eqnarray}
where $M_g(t)$ is the mass of gas in the system at time $t$, $(dM_g/dt)_{\rm in}$ is the 
infall rate of pristine gas, $\psi(t)$ is the star formation rate (SFR), $F_{\rm out}(t)$ is the 
rate of gas outflow, $M_{\rm Fe}(t)$ is the mass of Fe in the gas, and 
$P_{\rm Fe}(t)$ is the net rate of Fe production by all sources in the system.
We assume that the SFR is proportional to the mass of gas in the system 
with an astration rate constant $\lambda_*$,
\begin{equation}
\psi(t)=\lambda_*M_g(t).
\end{equation}
Given $(dM_g/dt)_{\rm in}$, $F_{\rm out}(t)$, and $P_{\rm Fe}(t)$, 
Eqs.~\ref{eq-dmgdt} and \ref{eq-dmfedt}
can be solved with the initial conditions $M_g(0)=0$ and $M_{\rm Fe}(0)=0$.

The MD of a system measures the numbers of stars formed in different metallicity intervals
that survive until the present time.
We use [Fe/H] to measure metallicity. As the mass fraction of H changes very little over the 
history of the universe, we take [Fe/H]~$=\log Z_{\rm Fe}$, where
\begin{equation}
Z_{\rm Fe}(t)\equiv\frac{M_{\rm Fe}(t)}{X_{\rm Fe}^\odot M_g(t)}.
\end{equation}
Here $X_{\rm Fe}^\odot$ is the mass fraction of Fe in the sun.
We assume that the initial mass function of SF does not change with time
and is of the Salpeter form. Then the number of stars formed per unit mass interval 
per unit time is related to the SFR as
\begin{equation}
\frac{d^2N}{dmdt}=\left[\frac{\psi(t)}{M_\odot}\right]
\frac{m^{-2.35}}{\int_{m_l}^{m_u}m^{-1.35}dm},
\end{equation}
where $m$ is the stellar mass in units of $M_\odot$
with $m_l$ and $m_u$ being the lower and upper limits, respectively.
We take $m_l=0.1$ and $m_u=100$. Assuming that $Z_{\rm Fe}(t)$ 
increases monotonically with time, we obtain the MD
\begin{eqnarray}
\frac{dN}{d{\rm [Fe/H]}}&=&
\frac{\int_{0.1}^{m_{\rm max}(t)}(d^2N/dmdt)dm}{d{\rm [Fe/H]}/dt}\nonumber\\
&=&\left(\frac{\lambda_*}{\log e}\right)
\frac{\int_{0.1}^{m_{\rm max}(t)}m^{-2.35}dm}{\int_{0.1}^{100}m^{-1.35}dm}
\nonumber\\
&\times&\left[\frac{M_g(t)}{M_\odot}\right]\frac{Z_{\rm Fe}(t)}{dZ_{\rm Fe}/dt},
\label{eq-dndfeh}
\end{eqnarray}
where $m_{\rm max}(t)$ is the maximum mass of those stars 
formed at time $t$ that survive until the present time.
There is little SF in dSphs at the present time (cf. \cite{evan}). 
We assume that SF ended at time $t_f$ in a system. 
Then the total number of stars in the system at the present time is
\begin{eqnarray}
N_{\rm tot}&=&\int_0^{t_f}\int_{0.1}^{m_{\rm max}(t)}\frac{d^2N}{dmdt}dmdt
\nonumber\\
&=&\frac{\lambda_*}{\int_{0.1}^{100}m^{-1.35}dm}
\int_0^{t_f}\frac{M_g(t)}{M_\odot}\int_{0.1}^{m_{\rm max}(t)}\frac{dmdt}{m^{2.35}}.
\label{eq-ntot}
\end{eqnarray}
The integral involving $m_{\rm max}(t)$ in Eqs.~\ref{eq-dndfeh} and
\ref{eq-ntot} increases only by 6\% when $m_{\rm max}(t)$ increases from 
0.8 to 100. As stars with $m=0.8$ have a lifetime approximately equal to
the age of the universe, we take $m_{\rm max}(t)=0.8$ in both these 
equations to obtain the normalized MD
\begin{equation}
\frac{1}{N_{\rm tot}}\frac{dN}{d{\rm [Fe/H]}}=\frac{1}{\log e}
\left[\frac{M_g(t)}{\int_0^{t_f}M_g(t)dt}\right]\frac{Z_{\rm Fe}(t)}{dZ_{\rm Fe}/dt}.
\label{eq-nmd}
\end{equation}

\section{The Model}
The key input for our model is the infall rate $(dM_g/dt)_{\rm in}$, 
the outflow rate $F_{\rm out}(t)$, and the net
Fe production rate $P_{\rm Fe}(t)$. The latter two rates are closely related to 
the occurrences of CCSNe and SNe Ia in a system.
CCSNe are associated with massive stars ($8<m\leq 100$)
that evolve rapidly. In contrast, SNe Ia require consideration of the evolution 
of binaries involving lower-mass stars with longer lifetimes.
In all our previous studies (e.g., \cite{qw04}), we considered 
that the evolution timescale for SNe Ia was $\sim 1$~Gyr using the lifetime
of stars with $m\sim2$. This meant that SNe Ia would not contribute to the Fe
inventory during early epochs and was in accord with the general approach
used by other workers. The consequence of this assumption is that the MD 
for a system must have two peaks due to the assumed late onset of SNe Ia 
(e.g., \cite{qw04}). However, of the eight dSphs studied in
\cite{kirby11a}, only a single peak is observed in the MD for Fornax, Leo I, 
Leo II, Sextans, Draco, and Canes Venatici I, and there is only some indication 
for two peaks in the MD for Ursa Minor and perhaps Sculptor.
From this we conclude that there must be a prompt component of SNe Ia
that start in a system on much shorter timescales than $\sim 1$~Gyr.
Such a component is supported by both supernova surveys 
(e.g., \cite{maoz11}) and models that considered detailed evolution of 
various binary configurations (e.g., \cite{gr83,greggio05}).
The occurrences of ``prompt'' and ``delayed'' SNe Ia were investigated 
in \cite{mann06}, where it was argued that both populations were present 
at high redshift. The nature of these two classes of SNe Ia remains unclear. 
For simplicity, we will lump the Fe production by these sources together with
that by CCSNe and ignore the time delay between the birth and death of 
all SN progenitors, except when the amount of gas in the system is so low
that SNe Ia would dominate. A detailed treatment of
SNe Ia that takes into account their evolution timescales will be discussed 
in a subsequent paper.

Under our assumption, the total rate of CCSNe and SNe Ia, and hence
$P_{\rm Fe}(t)$, are proportional to the SFR. We further assume that the 
rate of outflows driven by these SNe is also proportional to the SFR.
Specifically, we take
\begin{eqnarray}
F_{\rm out}(t)&=&\eta\lambda_*M_g(t),\label{eq-fout}\\
P_{\rm Fe}(t)&=&\lambda_{\rm Fe}X_{\rm Fe}^\odot M_g(t),\label{eq-pfe}
\end{eqnarray}
where $\eta$ is a dimensionless constant that measures the efficiency
of the SN-driven outflows and $\lambda_{\rm Fe}$ is a rate constant that
is proportional to $\lambda_*$ and the effective Fe yield of SNe. 
We take the infall rate to be
\begin{equation}
\left(\frac{dM_g}{dt}\right)_{\rm in}=\lambda_{\rm in}M_0
\frac{(\lambda_{\rm in}t)^\alpha}{\Gamma(\alpha+1)}\exp(-\lambda_{\rm in}t),
\label{eq-infall}
\end{equation}
where $\lambda_{\rm in}$ is a rate constant, $M_0$ is the total mass of gas infall over
$0\leq t<\infty$, and $\Gamma(\alpha+1)$ with $\alpha>-1$ is the Gamma function
of argument $\alpha+1$. 
The above modified exponential form was specifically 
chosen to explore the role of the time dependence of the infall rate in chemical 
evolution (cf. \cite{lb75}). 
The infall rate peaks at $t=0$ for $-1<\alpha\leq0$, and the peak time increases to
$\alpha/\lambda_{\rm in}$ for $\alpha>0$. The form with $\alpha>0$ allows a slow start of 
significant gas accumulation in the system.

With the above assumptions,
Eqs.~\ref{eq-dmgdt} and \ref{eq-dmfedt} become
\begin{eqnarray}
\frac{dM_g}{dt}&=&\lambda_{\rm in}M_0
\frac{(\lambda_{\rm in}t)^\alpha}{\Gamma(\alpha+1)}
\exp(-\lambda_{\rm in}t)-\lambda M_g(t),\label{eq-dmgdt1}\\
\frac{dM_{\rm Fe}}{dt}&=&\lambda_{\rm Fe}X_{\rm Fe}^\odot M_g(t)
-\lambda M_{\rm Fe}(t),\label{eq-dmfedt1}
\end{eqnarray}
where $\lambda\equiv(1+\eta)\lambda_*$. 
Note that for $\lambda_{\rm in}\ll\lambda$,
the solutions to the above equations approach a secular state for which 
$M_g(t)\approx(dM_g/dt)_{\rm in}/\lambda$ and
$Z_{\rm Fe}\approx\lambda_{\rm Fe}/\lambda$.
This is analogous to the quasi-steady state for the metallicity of the 
interstellar medium first proposed in \cite{lars72}.
For simplicity, we assume
that $\lambda_{\rm in}$ and $\lambda$ are so large that $t_f=\infty$ 
can be used effectively in Eq.~\ref{eq-nmd} for the MD. 
As $M_g(0)=M_g(\infty)=0$, integrating Eq.~\ref{eq-dmgdt1}
over $t$ gives $\lambda\int_0^\infty M_g(t)dt=M_0$. So the total
gas mass used in SF is
$\lambda_*\int_0^\infty M_g(t)dt=M_0/(1+\eta)$ and the remainder
of the gas infall 
is blown out as outflows. We assume that all outflows are lost from 
the system into the broader intergalactic medium (IGM), 
thus enriching the latter in metals. 
This approach gives a natural cutoff to the chemical evolution of
the system when the total mass of gas lost to SF and outflows is 
equal to the total mass of gas infall (partial recycling of outflows 
would increase the degree of chemical enrichment
but is ignored here for simplicity). At the present time,
the total mass of stars in the system can be estimated as
\begin{equation}
M_*=\frac{\int_{0.1}^{0.8}m^{-1.35}dm}{\int_{0.1}^{100}m^{-1.35}dm}
\frac{M_0}{1+\eta}=9.65\times10^{-2}\left(\frac{M_h}{1+\eta}\right),
\label{eq-ms}
\end{equation}
where we have used $M_0=(\Omega_b/\Omega_m)M_h=0.17M_h$.
Here $M_h$ is the total mass inside the dark matter halo hosting
the system, and $\Omega_b$ and $\Omega_m$ are the fractional
contributions to the critical density of the universe from baryonic and
all matter, respectively.

\section{Exploration of Results from the Model}
The parameters governing the solutions to Eqs.~\ref{eq-dmgdt1} 
and \ref{eq-dmfedt1} are $\alpha$, $\lambda_{\rm in}$, $\lambda$, 
and $\lambda_{\rm Fe}$. We illustrate the dependences of the MD
on these parameters in the following subsections.

\subsection{Dependence of the MD on $\lambda_{\rm in}/\lambda$ 
with $\alpha=0$}
For $\alpha=0$, the solutions to Eqs.~\ref{eq-dmgdt1} and 
\ref{eq-dmfedt1} are
\begin{eqnarray}
M_g(t)&=&\frac{\lambda_{\rm in}}{\lambda_{\rm in}-\lambda}
M_0[\exp(-\lambda t)-\exp(-\lambda_{\rm in}t)],\label{eq-mg}\\
M_{\rm Fe}(t)&=&\frac{\lambda_{\rm Fe}\lambda_{\rm in}}
{(\lambda_{\rm in}-\lambda)^2}X_{\rm Fe}^\odot M_0
\{\exp(-\lambda_{\rm in}t)\nonumber\\
&+&[(\lambda_{\rm in}-\lambda)t-1]\exp(-\lambda t)\},
\end{eqnarray}
from which we obtain
\begin{eqnarray}
Z_{\rm Fe}(t)&=&\frac{\lambda_{\rm Fe}}{\lambda_{\rm in}-\lambda}
\left[\frac{(\lambda_{\rm in}-\lambda)t}{1-\exp[-(\lambda_{\rm in}-\lambda)t]}
-1\right],\label{eq-zfe}\\
\frac{1}{N_{\rm tot}}\frac{dN}{d{\rm [Fe/H]}}&=&
\frac{1}{\log e}\frac{\lambda_{\rm in}\lambda}{(\lambda_{\rm in}-\lambda)^2}
\frac{\{1-\exp[-(\lambda_{\rm in}-\lambda)t]\}^2}{\exp(\lambda t)}\nonumber\\
&\times&\frac{(\lambda_{\rm in}-\lambda)t-1+\exp[-(\lambda_{\rm in}-\lambda)t]}
{1-[1+(\lambda_{\rm in}-\lambda)t]\exp[-(\lambda_{\rm in}-\lambda)t]}.\label{eq-md}
\end{eqnarray}
For $\lambda_{\rm in}/\lambda=1$, the above two equations reduce to
\begin{eqnarray}
Z_{\rm Fe}(t)&=&\frac{\lambda_{\rm Fe}t}{2},\\
\frac{1}{N_{\rm tot}}\frac{dN}{d{\rm [Fe/H]}}&=&
\frac{(\lambda t)^2}{\log e}\exp(-\lambda t).
\end{eqnarray}
In the limit $\lambda_{\rm in}=\infty$,
which is equivalent to setting $M_g(0)=M_0$ and $(dM_g/dt)_{\rm in}=0$ 
for $t>0$, the results are
\begin{eqnarray}
Z_{\rm Fe}(t)&=&\lambda_{\rm Fe}t,\label{eq-zfelim}\\
\frac{1}{N_{\rm tot}}\frac{dN}{d{\rm [Fe/H]}}&=&
\frac{\lambda t}{\log e}\exp(-\lambda t).
\label{eq-mdlim}
\end{eqnarray} 

\begin{figure}[ht]
\centerline{\includegraphics[angle=270,scale=.3]{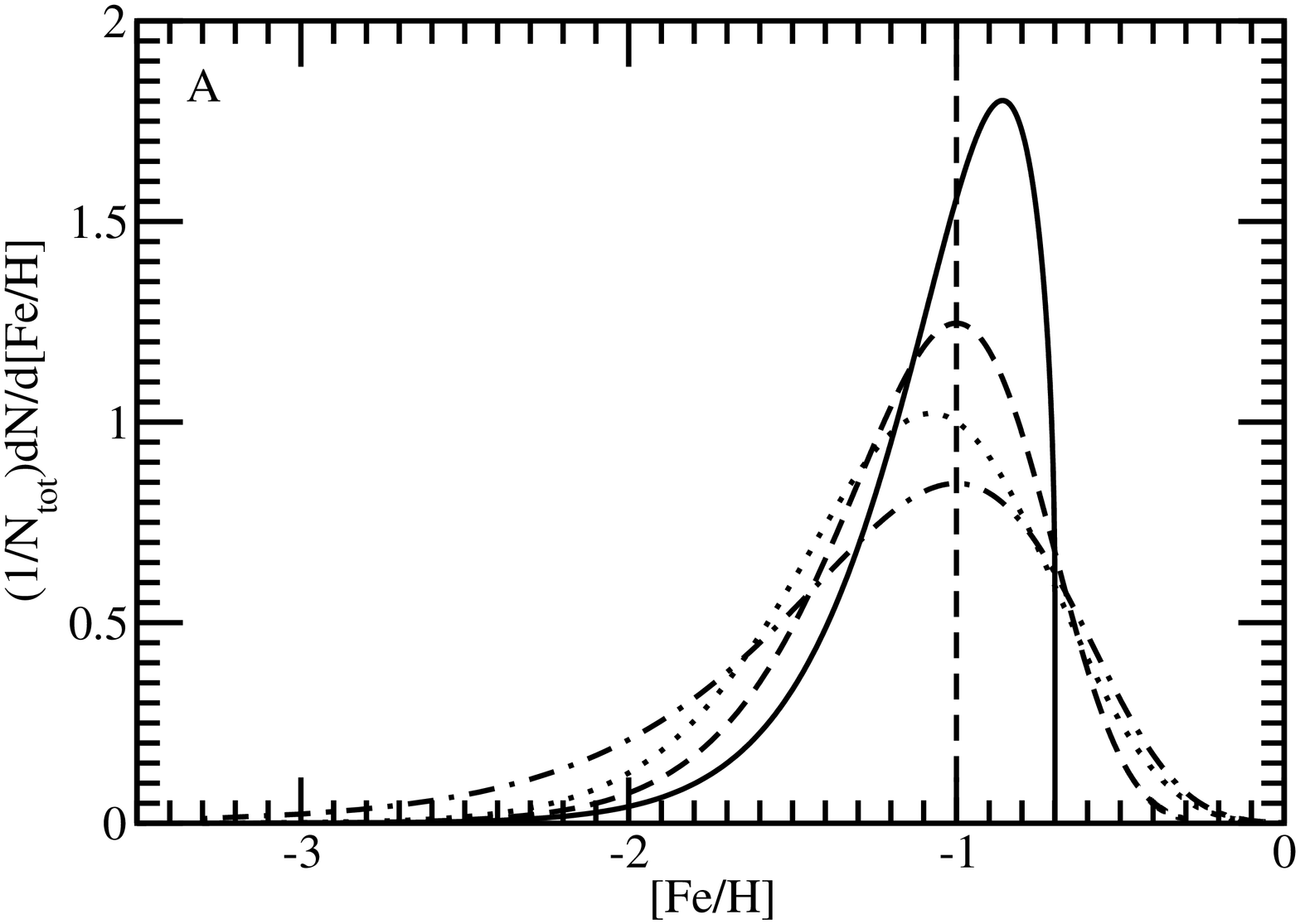}}
\vspace{-0.45cm}
\centerline{\includegraphics[angle=270,scale=.3]{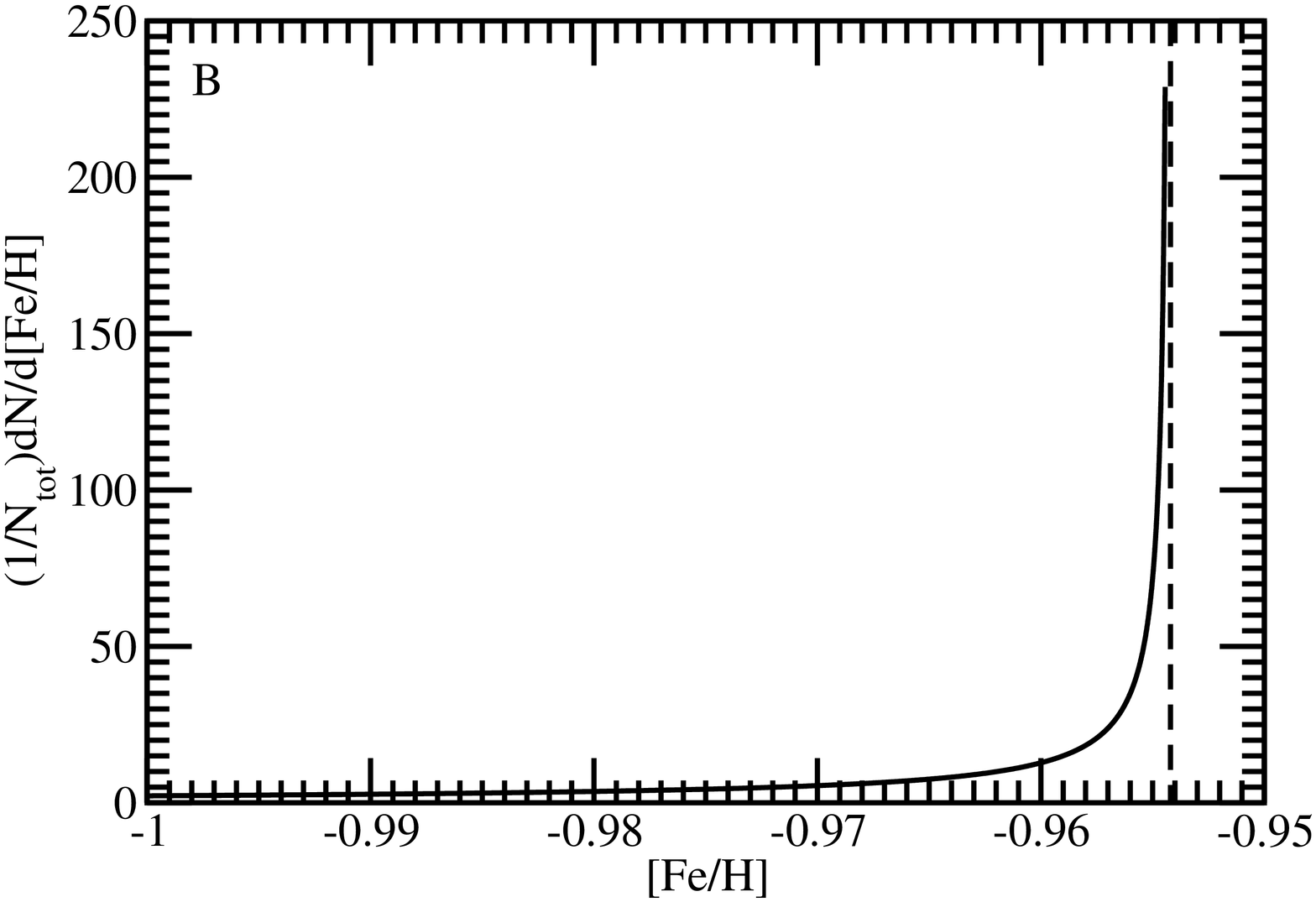}}
\caption{Example MDs for $\lambda_{\rm Fe}/\lambda=0.1$
and various values of $\lambda_{\rm in}/\lambda$ with $\alpha=0$.
Note that in general the shape of the MD is determined by 
$\lambda_{\rm in}/\lambda$ and changing $\lambda_{\rm Fe}/\lambda$
only translates the MD along the [Fe/H]-axis. 
(A)  The solid, dashed, dotted, and dot-dashed curves are the MDs for 
$\lambda_{\rm in}/\lambda=1/2$, 1, 2, and $\infty$, respectively.
The vertical dashed line indicates the peak at 
[Fe/H]~$=\log(\lambda_{\rm Fe}/\lambda)=-1$ 
for the dashed and dot-dashed curves. (B) The solid curve is the
MD for $\lambda_{\rm in}/\lambda=0.1$ and the dashed line indicates 
the limiting value of 
[Fe/H]~$=\log[\lambda_{\rm Fe}/(\lambda-\lambda_{\rm in})]=-0.954$. 
Concentration of stars in an extremely narrow range of [Fe/H] 
as shown by the solid curve 
is typically not observed for dSphs but strongly
resembles what is observed for most globular clusters.}
\label{fig-mda0}
\end{figure}

In general, Eqs.~\ref{eq-zfe} and \ref{eq-md} give the MD 
as a function of [Fe/H] in parametric form. This MD only depends on 
$\lambda_{\rm in}/\lambda$ and $\lambda_{\rm Fe}/\lambda$, but not on
the absolute values of these rates (this is true so long as $\lambda t_f\gg 1$ 
and $\lambda_{\rm in}t_f\gg 1$, see Eqs.~\ref{eq-nmd} and \ref{eq-mg}).
For a fixed $\lambda_{\rm in}/\lambda$, changing $\lambda_{\rm Fe}/\lambda$
only translates the MD along the [Fe/H]-axis. This can be most easily seen 
in the special cases of $\lambda_{\rm in}/\lambda=1$ and 
$\lambda_{\rm in}=\infty$, where $N_{\rm tot}^{-1}dN/d{\rm [Fe/H]}$ is a simple
function of $\lambda t$ and [Fe/H] differs from $\log(\lambda t)$ only by a
shift of $\log(\lambda_{\rm Fe}/2\lambda)$ and $\log(\lambda_{\rm Fe}/\lambda)$,
respectively. The shape of the MD is determined by 
$\lambda_{\rm in}/\lambda$. This can be seen from Fig.~\ref{fig-mda0}A,
which uses $\lambda_{\rm Fe}/\lambda=0.1$ and
shows that as $\lambda_{\rm in}/\lambda$ increases from 1/2 to $\infty$, 
the position of the peak of the MD changes slightly [but staying close to 
${\rm [Fe/H]}=\log(\lambda_{\rm Fe}/\lambda)=-1$, the exact peak
position for $\lambda_{\rm in}/\lambda=1$ and $\lambda_{\rm in}=\infty$]
and the shape of the MD becomes broader.
Note the sharp cutoff of the MD to the right of the peak for 
$\lambda_{\rm in}/\lambda=1/2$ with no stars formed above
[Fe/H]~$=\log[\lambda_{\rm Fe}/(\lambda-\lambda_{\rm in})]=-0.7$.

The case of $\lambda_{\rm in}/\lambda<1/2$ requires separate 
discussion. For illustration, we again take
$\lambda_{\rm Fe}/\lambda=0.1$ and show the MD for
$\lambda_{\rm in}/\lambda=0.1$ in Fig.~\ref{fig-mda0}B.
This MD has an extremely sharp peak with 90\% of the stars having 
$-1.154\leq{\rm [Fe/H]}<-0.954$. This is simply a case close to 
secular equilibrium (see discussion below Eqs.~\ref{eq-dmgdt1}
and \ref{eq-dmfedt1}; cf. \cite{lars72}). Piling up of stars in an
extremely narrow metallicity range is typically not observed for 
dwarf galaxies but strongly resembles what is observed for most 
globular clusters. In general, for $\lambda_{\rm in}/\lambda$ 
significantly below 1/2, stars are concentrated immediately below 
[Fe/H]~$=\log[\lambda_{\rm Fe}/(\lambda-\lambda_{\rm in})]$.
For comparison, the solid curve in Fig.~\ref{fig-mda0}A shows that
the MD for $\lambda_{\rm in}/\lambda=1/2$
first rises to a peak and then sharply drops to zero as 
[Fe/H]~$\to\log[\lambda_{\rm Fe}/(\lambda-\lambda_{\rm in})]$.
This general behavior also applies to $\lambda_{\rm in}/\lambda>1/2$ 
but with a more extended tail at high metallicities for a larger 
$\lambda_{\rm in}/\lambda$ (see Fig.~\ref{fig-mda0}A).

\subsection{Dependence of the MD on $\alpha$ with $\lambda_{\rm in}=\lambda$}
Based on the above discussion and the example MDs
shown in Fig.~\ref{fig-mda0}, it seems reasonable to 
choose $\lambda_{\rm in}=\lambda$ and explore possible MDs 
for different values of $\alpha$ and $\lambda_{\rm Fe}/\lambda$. 
For $\lambda_{\rm in}=\lambda$
and $\alpha>-1$, the solutions to Eqs.~\ref{eq-dmgdt1} and 
\ref{eq-dmfedt1} are
\begin{eqnarray}
M_g(t)&=&M_0\frac{(\lambda t)^{\alpha+1}}{\Gamma(\alpha+2)}\exp(-\lambda t),\\
M_{\rm Fe}(t)&=&\frac{\lambda_{\rm Fe}}{\lambda}X_{\rm Fe}^\odot M_0
\frac{(\lambda t)^{\alpha+2}}{\Gamma(\alpha+3)}\exp(-\lambda t),
\end{eqnarray}
which give
\begin{eqnarray}
Z_{\rm Fe}(t)&=&\frac{\lambda_{\rm Fe}t}{\alpha+2},\label{eq-zfea}\\
\frac{1}{N_{\rm tot}}\frac{dN}{d{\rm [Fe/H]}}&=&\frac{1}{\log e}
\frac{(\lambda t)^{\alpha+2}}{\Gamma(\alpha+2)}\exp(-\lambda t).\label{eq-mda}
\end{eqnarray}
The above MD again has a peak at [Fe/H]~$=\log(\lambda_{\rm Fe}/\lambda)$
for all $\alpha>-1$. Changing $\lambda_{\rm Fe}/\lambda$ 
only translates the MD, shifting the peak in particular, along the [Fe/H]-axis.
The shape of the MD is determined by $\alpha$.
For a specific $\lambda_{\rm Fe}/\lambda$,
the MD becomes narrower while peaking at the same [Fe/H]
when $\alpha$ increases above $-1$.
A positive $\alpha$ reflects slower infall at the start of the system,
which suppresses SF at the early stages and 
reduces the extent of the low-metallicity tail of the MD. 
For $\alpha< 0$, the initial infall rate is enhanced and consequently,
more stars are formed at lower metallicities.

Note that if we take the limit $\alpha=-1$, the MD given by
Eqs.~\ref{eq-zfea} and \ref{eq-mda}
coincides with that for $\lambda_{\rm in}=\infty$ and $\alpha=0$
(see Eqs.~\ref{eq-zfelim} and \ref{eq-mdlim}). Thus, for
$\lambda_{\rm Fe}/\lambda=0.1$, as $\alpha$ increases from $-1$ to 0, 
the MD for $\lambda_{\rm in}=\lambda$ changes from the dot-dashed
to the dashed curve shown in Fig.~\ref{fig-mda0}A while peaking at the
same [Fe/H]~$=-1$. More example MDs for $\lambda_{\rm in}=\lambda$
and $\alpha\geq-1$ are shown in Fig.~\ref{fig-obs} 
where we compare them with observations of dSphs.

\section{Comparison with Observations}
We now explore the implications of our model for observations of dSphs.
We first discuss the MDs using the high-quality data set of \cite{kirby11a} and 
then study the relationship between the mean metallicity 
$\langle{\rm [Fe/H]}\rangle$ and the stellar mass $M_*$ of dSphs.

\subsection{MDs for dSphs}
\begin{figure*}
\centerline{\includegraphics[angle=270,scale=.22]{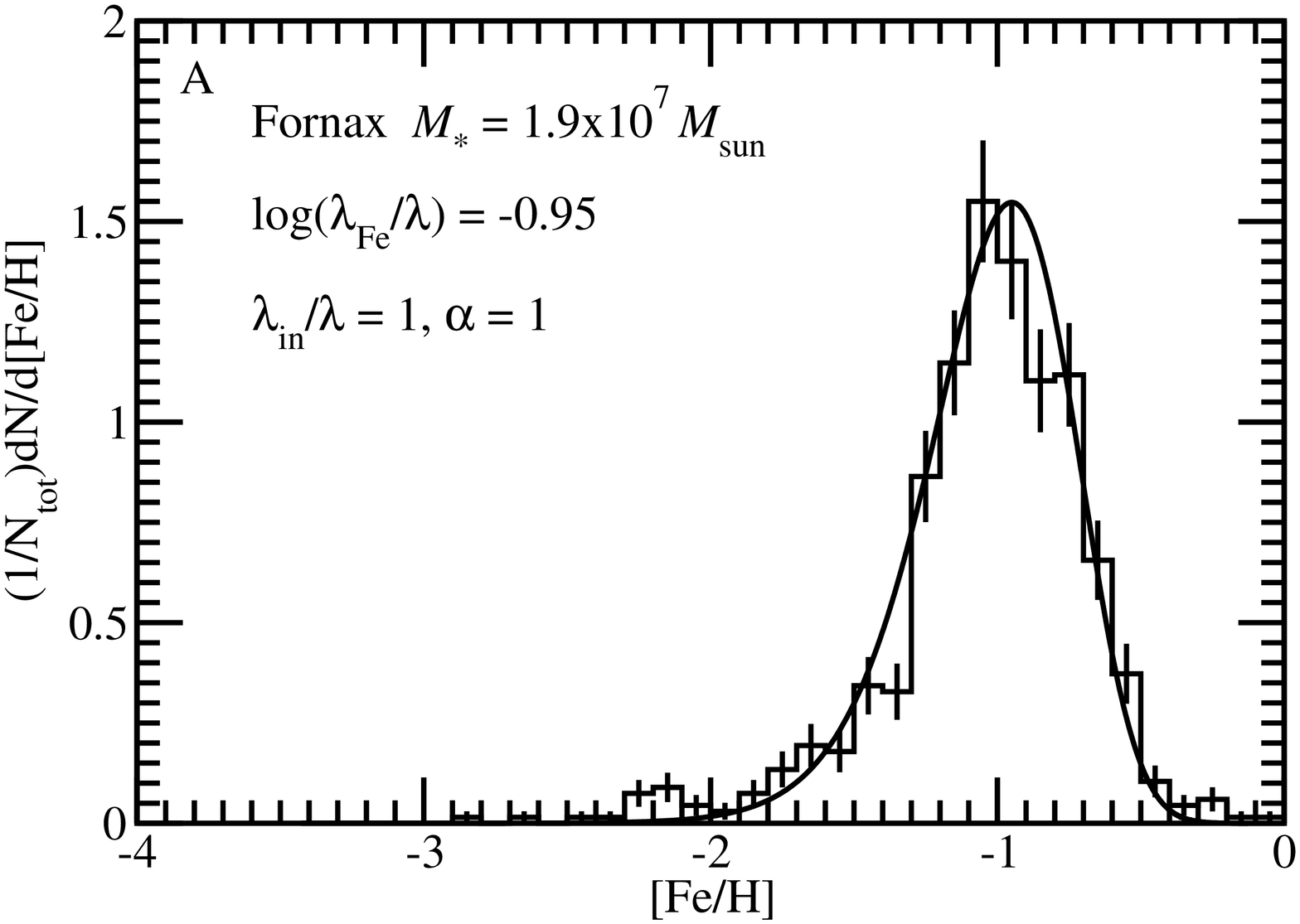}%
\includegraphics[angle=270,scale=.22]{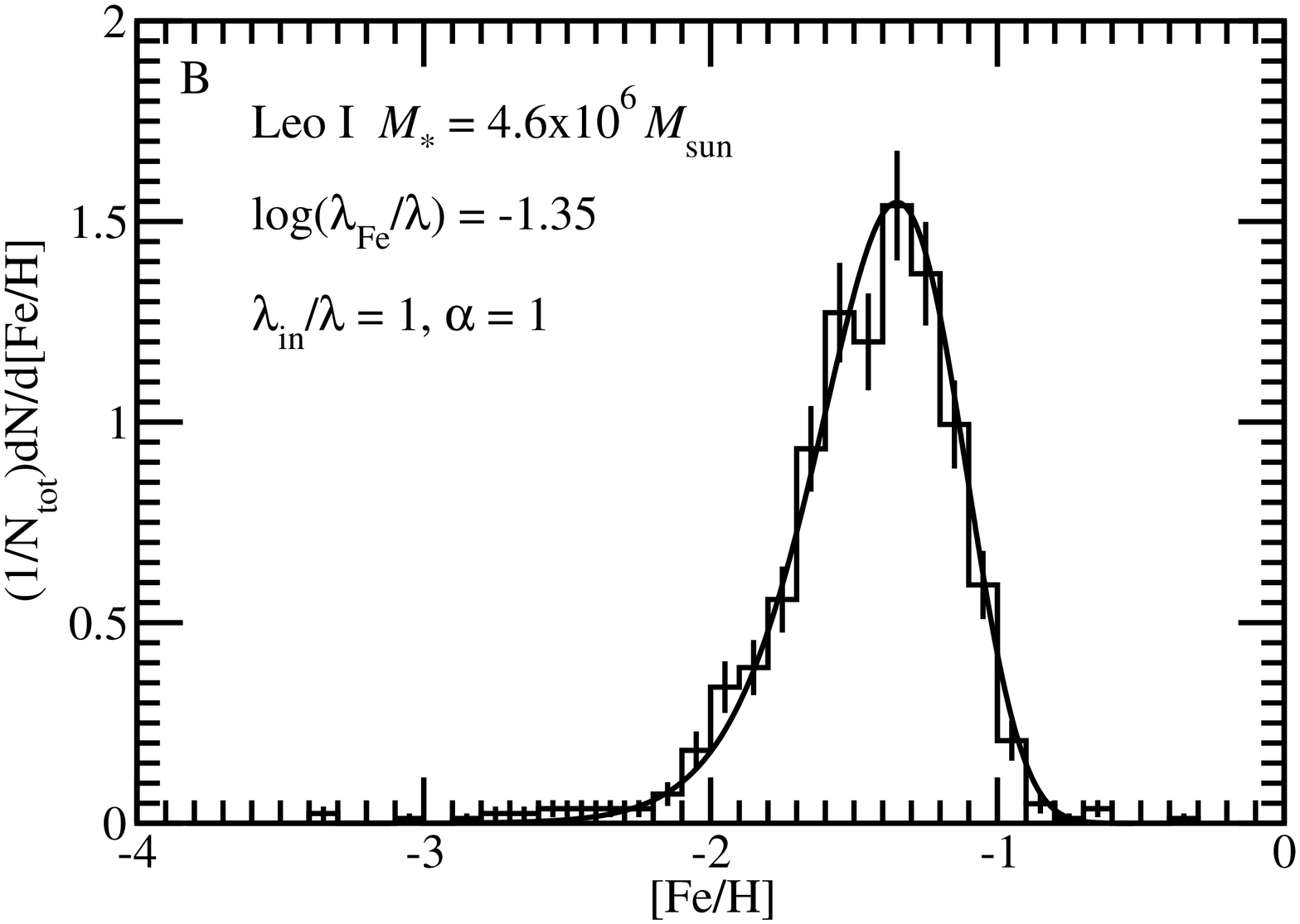}%
\includegraphics[angle=270,scale=.22]{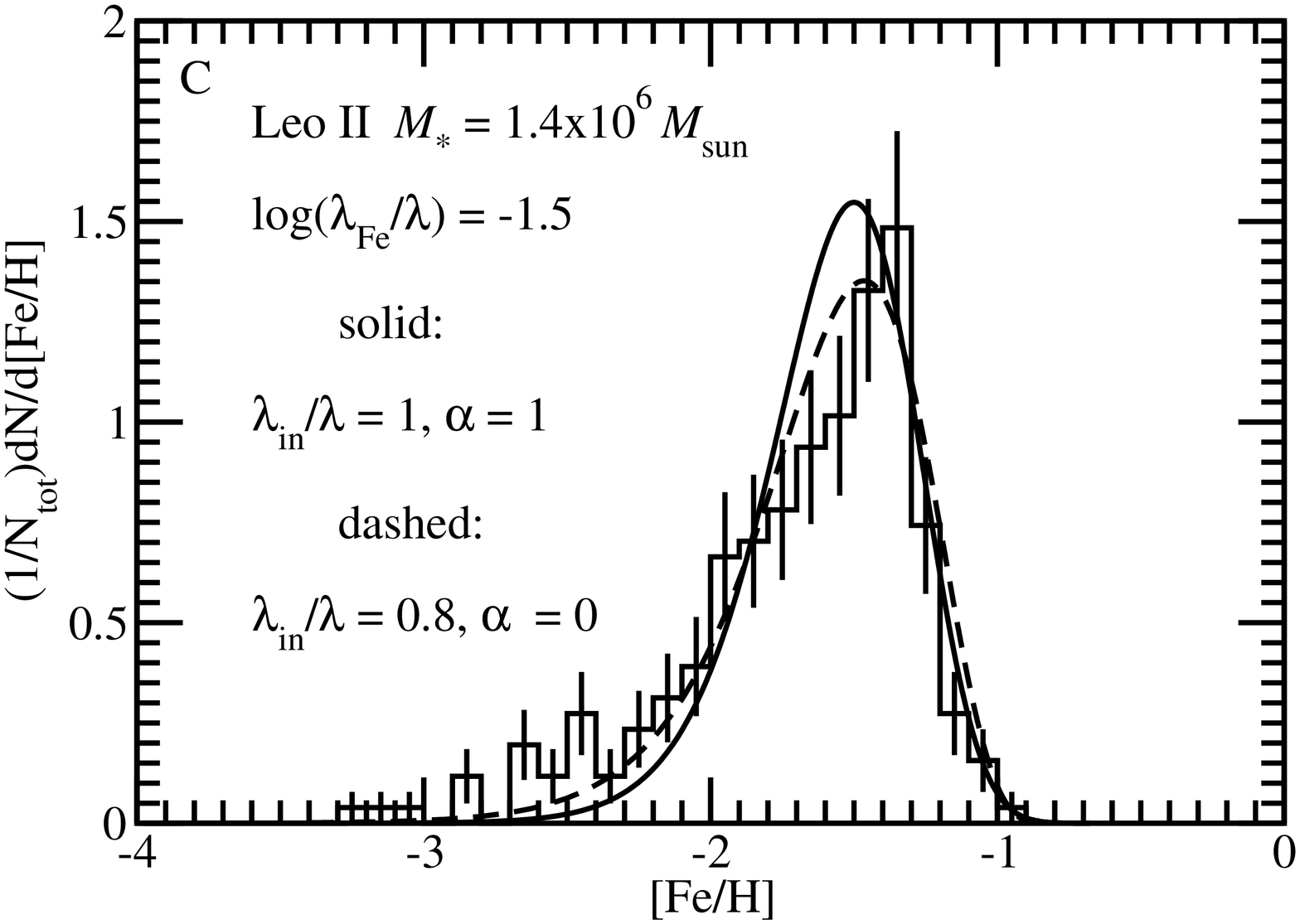}}
\vspace{-0.3cm}
\centerline{\includegraphics[angle=270,scale=.22]{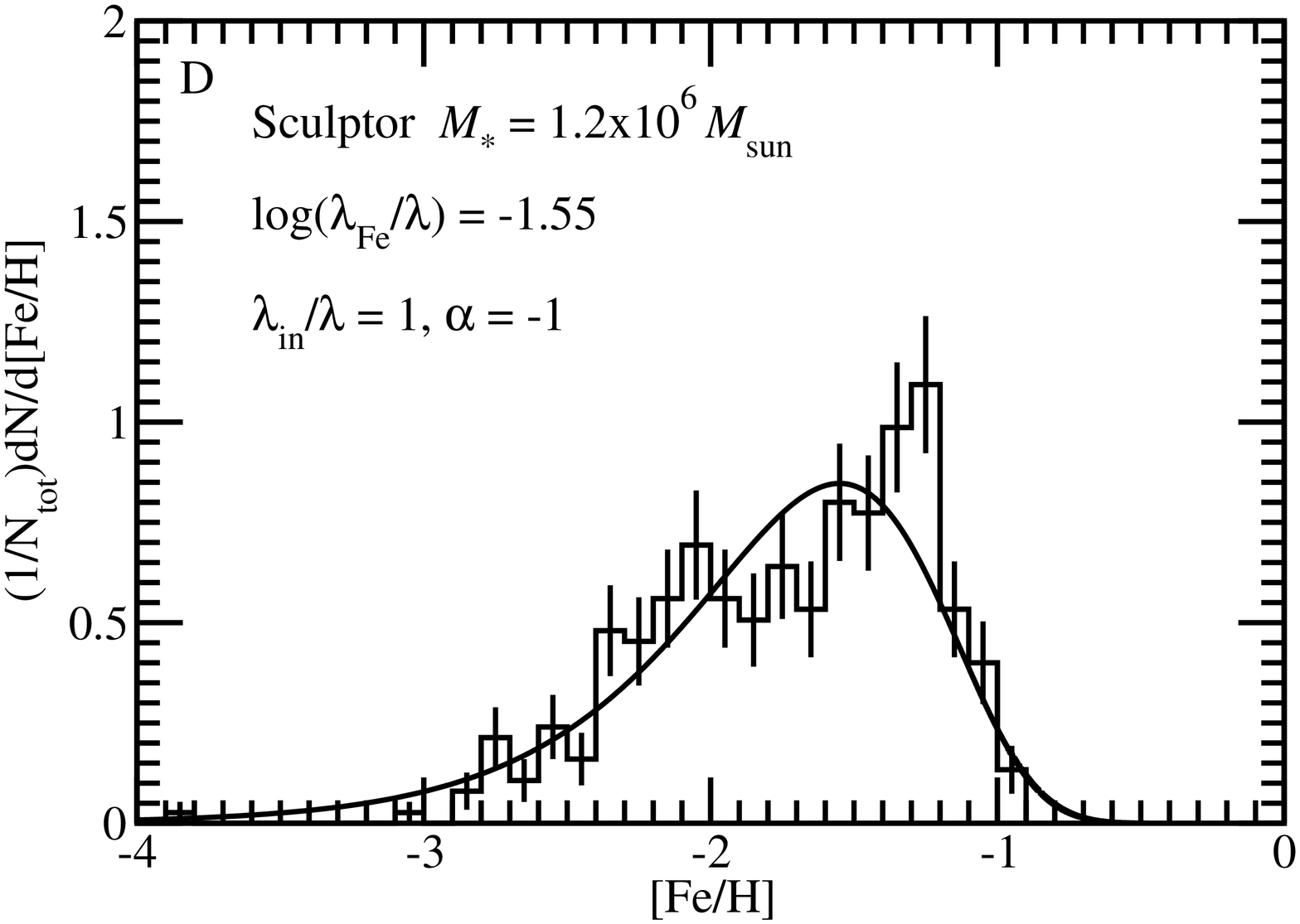}%
\includegraphics[angle=270,scale=.22]{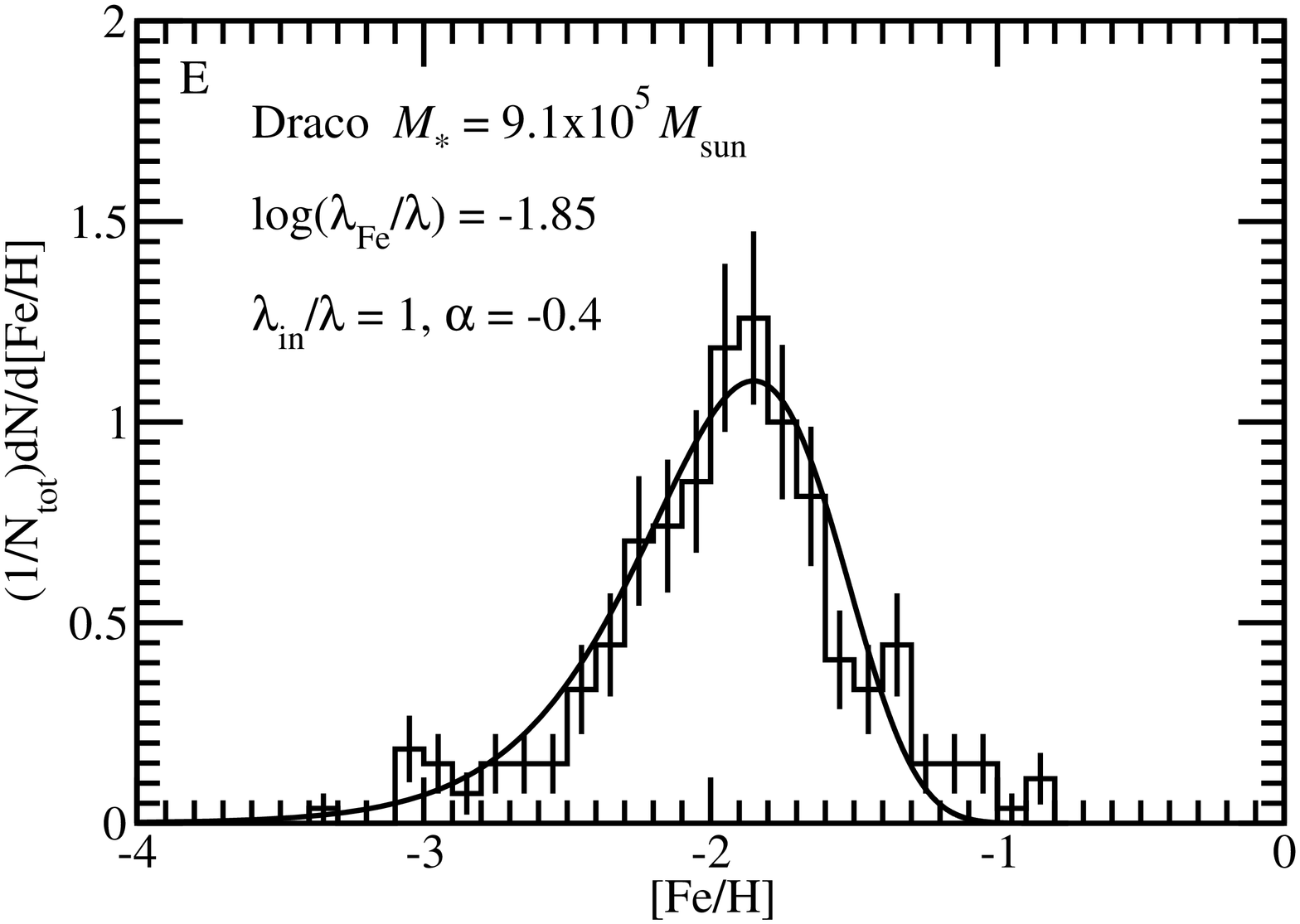}%
\includegraphics[angle=270,scale=.22]{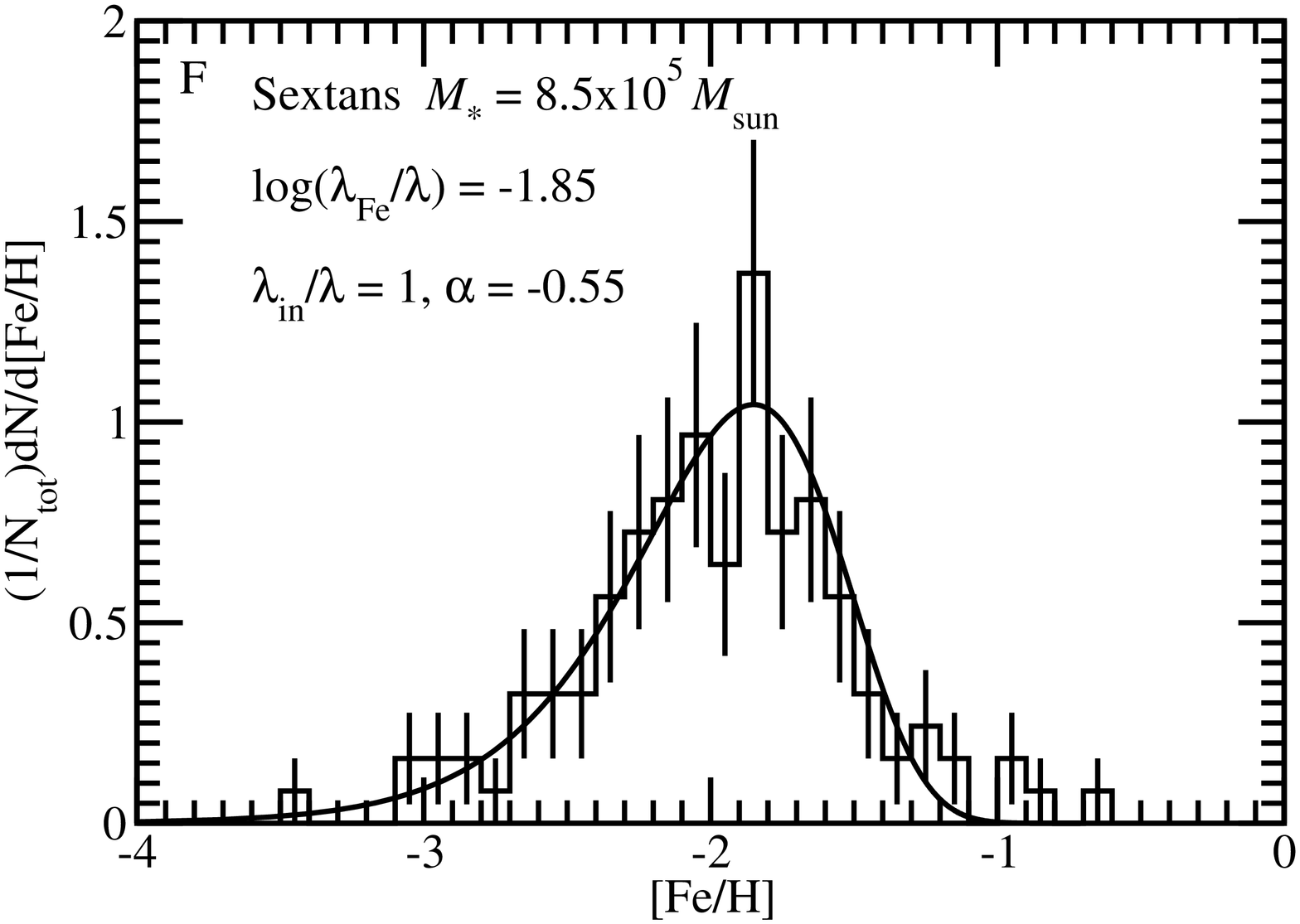}}
\vspace{-0.3cm}
\centerline{\includegraphics[angle=270,scale=.22]{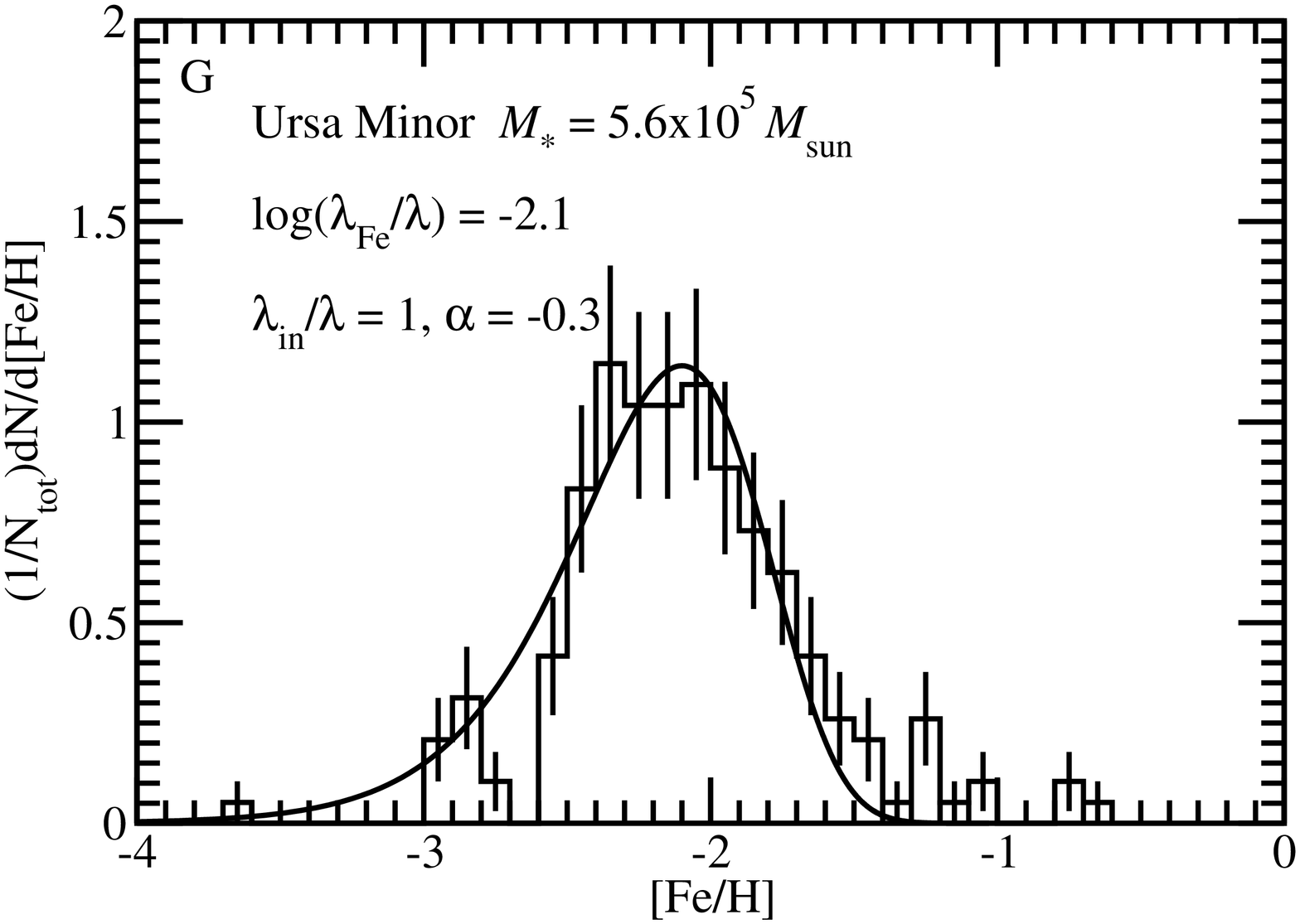}%
\includegraphics[angle=270,scale=.22]{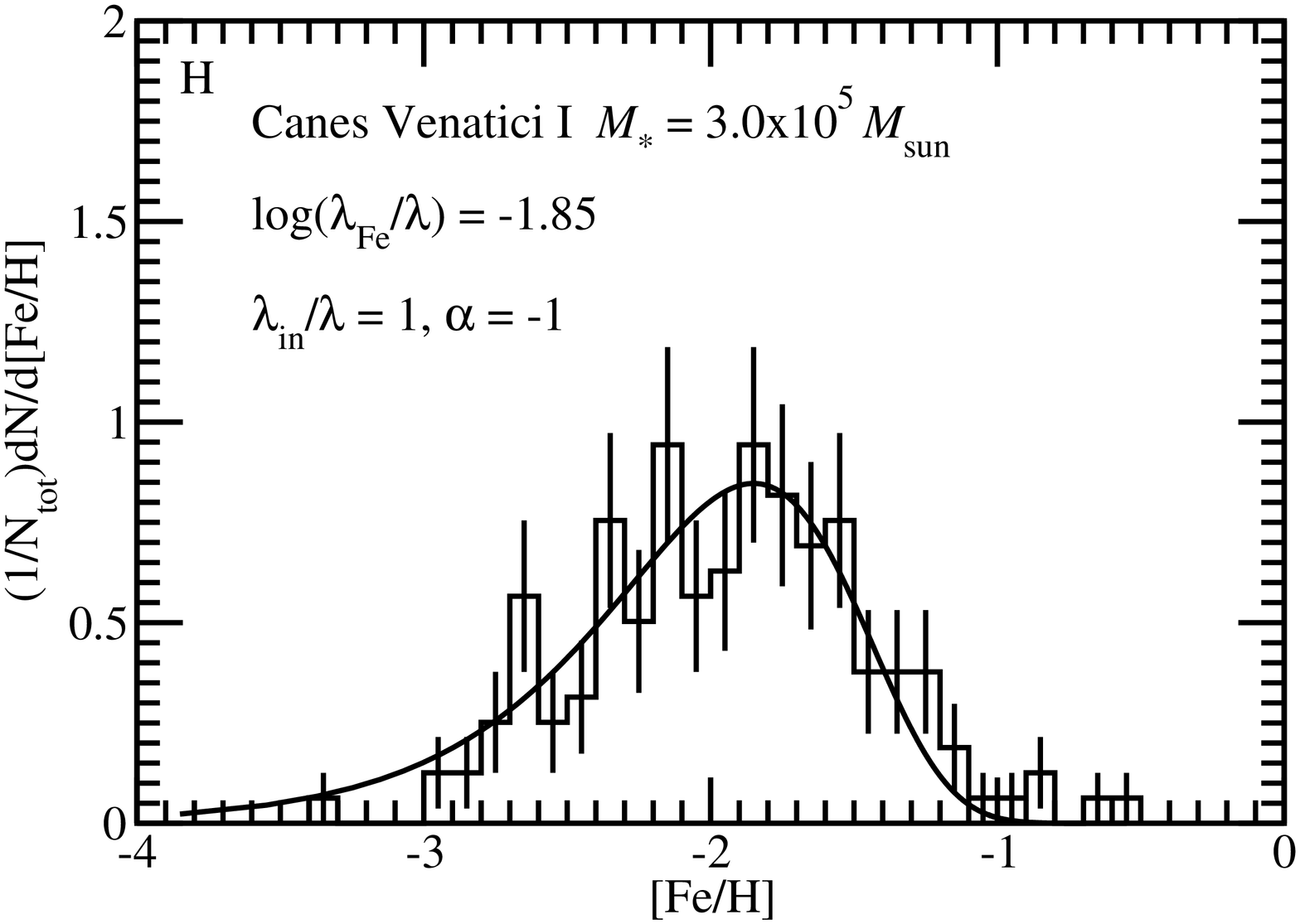}}
\caption{Comparison of model MDs with observations of dSphs.
The data are taken from \cite{kirby11a} and shown as histograms with
error bars. The model MDs assume the indicated parameters and
are shown as curves. The dashed curve in (C) provides a better fit to 
the data than the solid curve. Values of $M_*$ are taken from 
\cite{woo} for (A)--(G) and from \cite{martin} for (H).
Note that if the baryonic matter is not always blown out of the dark 
matter halo but returns to the gas mass in a dSph
after some time, then the curves will have a leading-edge tail 
going to higher [Fe/H].}
\label{fig-obs}
\end{figure*}

Important medium-resolution data on the MDs for dSphs
were provided in \cite{kirby11a} and are summarized in 
Fig.~\ref{fig-obs}. Careful inspection of these data shows that there is 
only some indication for two peaks in the MD for Ursa Minor and perhaps
Sculptor. An MD with two peaks 
would be typical if the turn-on of SNe Ia were sudden and
with a significant delay relative to CCSNe. It was the lack of such MDs from
observations that led us to pursue a model where the net Fe production rate 
is without discontinuities.

We focus on models with $\lambda_{\rm in}=\lambda$ and 
different values of $\alpha$. In this case, a model MD is specified by
$\lambda_{\rm Fe}/\lambda$ and $\alpha$, which determine its peak
position [Fe/H]~$=\log(\lambda_{\rm Fe}/\lambda)$ and
its shape, respectively. 
Positive values of $\alpha$ corresponding 
to slower initial infall give rise to narrower MDs while negative values
corresponding to more rapid initial infall result in more extended MDs.
As the MD is normalized, its peak height 
can be used to estimate $\alpha$ effectively.
Using the position and the height of the peak for the observed MD
as guides, we fit a model MD for each of the eight dSphs
reported in \cite{kirby11a}. The observed and fitted MDs are shown
as histograms and curves, respectively, in Fig.~\ref{fig-obs}.
The adopted values of $\lambda_{\rm Fe}/\lambda$ and $\alpha$ are
indicated for each galaxy. These values were not obtained from
the best fits, but were simply picked to illustrate the overall 
adequacy of our model. Very good fits are obtained for 
Fornax and Leo I, which are the most massive of the eight dSphs.
The fits for the least 
massive four, Draco, Sextans, Ursa Minor, and Canes Venatici I,
are rather good, although the model MDs appear to underestimate 
their stellar populations at the highest metallicities. The only exceptions 
are Leo II and Sculptor with intermediate $M_*$.
Using the same $\lambda_{\rm Fe}/\lambda$, we obtained a better
fit to the data for Leo II with $\lambda_{\rm in}/\lambda=0.8$ and 
$\alpha=0$ (dashed curve in Fig.~\ref{fig-obs}C) than with 
$\lambda_{\rm in}/\lambda=1$ and $\alpha=1$
(solid curve). However, neither type of MD
can provide a good fit to the data
for Sculptor, which may represent an MD with two peaks.
There is perhaps a more clear indication for such an MD
for Ursa Minor. We note that if infall took a more
complicated form than Eq.~\ref{eq-infall}, then 
the shape of the MD would change accordingly. It is possible 
that the broadened peak in the MD for Sculptor might be 
explained by an increase in the infall rate after the assumed
smooth rate in Eq.~\ref{eq-infall} peaked. However,
in no case can a significant part of the enriched outflows be 
returned to the infalling matter as this would produce a 
population of stars with high [Fe/H] values, which is not observed.

In a previous study \cite{kirby11a} with alternative modelling, 
detailed multi-parameter fitting was used.
As can be seen from Fig.~\ref{fig-obs}, a good to 
excellent fit of the model to the data can be obtained in our 
approach. The model thus appears to give a good description of 
the MDs of dSphs and is easily understood 
in terms of physical processes.

\subsection{Relationship between $\langle{\rm [Fe/H]}\rangle$ and 
$M_*$ for dSphs}
\begin{figure}
\centerline{\includegraphics[angle=270,scale=.3]{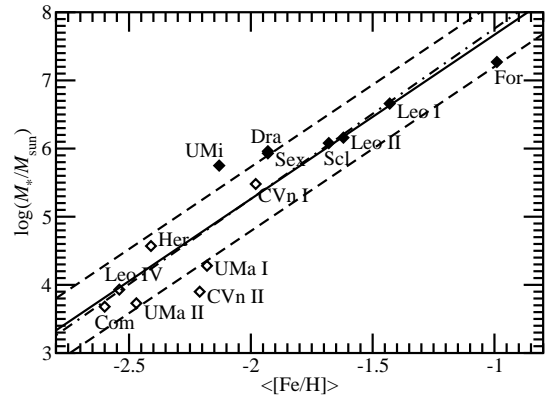}}
\caption{The relationship between $\log(M_*/M_\odot)$ and
$\langle{\rm [Fe/H]}\rangle$ for dSphs. Values of
$M_*$ for filled and open diamonds are taken from
\cite{woo} and \cite{martin}, respectively. All values of
$\langle{\rm [Fe/H]}\rangle$ used here are taken from
\cite{kirby11a}. The solid line is the least-square fit to the data
and the dashed lines indicate the $1\sigma$ error in
$\log(M_*/M_\odot)$. The dot-dashed line has a slope of 2.5
and is nearly the same as the solid line.}
\label{fig-fems}
\end{figure}

The observed MDs of dSphs require that 
$\lambda_{\rm Fe}/\lambda\sim 0.01$--0.1 (see Fig.~\ref{fig-obs}). 
As $\lambda_{\rm Fe}/\lambda\sim (1+\eta)^{-1}$
(see derivation of $\lambda_{\rm Fe}$ in SI Text),
this requires that $\eta\sim 10$--100.
Consequently, only a fraction $(1+\eta)^{-1}\sim 1$--10\% of the gas 
falling into the dark matter halo hosting a dSph is used in SF 
and the rest is blown out of the halo by SN-driven outflows. 
The MD of a dSph peaks at essentially its mean metallicity
$\langle{\rm [Fe/H]}\rangle\approx\log(\lambda_{\rm Fe}/\lambda)\sim-\log\eta$.
The lower $\langle{\rm [Fe/H]}\rangle$ a dSph has, 
the lower fraction of the infalling gas is stored in its stars.
There is thus a direct relationship between $\langle{\rm [Fe/H]}\rangle$ 
and the stellar mass $M_*$ for dSphs. Using the data on 
$\langle{\rm [Fe/H]}\rangle$ in \cite{kirby11a} and those
on $M_*$ in \cite{woo,martin}, we show this relationship in
Fig.~\ref{fig-fems}. The dot-dashed line in 
Fig.~\ref{fig-fems} has a slope of 2.5 and passes through the
point defined by the average values of the data. It is nearly 
the same as the least-square fit (solid line) and is in excellent agreement
with the result in \cite{woo} (see SI Text).

As discussed in \cite{dekel}, the slope of 2.5 for the
relationship between $\log(M_*/M_\odot)$ and $\langle{\rm [Fe/H]}\rangle$
has a simple physical explanation.
In general, this slope follows from the relationship between
the radius $r_h$ and the total mass $M_h$ of the dark matter halo 
hosting a dSph in addition to the dependences of $M_*$ and
$\langle{\rm [Fe/H]}\rangle$ on $\eta$ discussed above.
The outflow efficiency $\eta$ is inversely proportional to the depth of 
the gravitational potential well of the dark matter halo,
\begin{equation}
\eta\propto r_h/M_h\propto M_h^{\beta-1},
\end{equation}
where we have assumed $r_h\propto M_h^\beta$ in the second
step. This gives
\begin{equation}
\langle{\rm [Fe/H]}\rangle\sim-\log\eta\sim(1-\beta)\log M_h+\mbox{const.}
\end{equation} 
In addition, Eq.~\ref{eq-ms} gives
\begin{equation}
M_*\propto M_h/\eta\propto M_h^{2-\beta}.
\end{equation}
Thus, a slope of 2.5 for the relationship between $\log(M_*/M_\odot)$ 
and $\langle{\rm [Fe/H]}\rangle$ requires $(2-\beta)/(1-\beta)=2.5$,
or $\beta=1/3$, which is in agreement with the framework of 
hierarchical structure formation (e.g., Eq.~24 in \cite{loeb}).

\section{Conclusions}
We have shown that the MDs of dSphs can be reasonably well 
described quantitatively by a phenomenological model based on general
considerations of gas infall into a dark matter halo, astration in the 
condensed gas, and SN-driven gas outflows. The peak position of the MD
is governed by the ratio of the rate constants for net Fe production and
net gas loss to astration and outflows,
$\lambda_{\rm Fe}/\lambda=\lambda_{\rm Fe}/[(1+\eta)\lambda_*]$.
The observations require high efficiency $\eta\sim 10$--360 
(see Table~S1) for SN-driven
outflows and hence, massive gas loss from dark matter halos associated 
with dSphs. The model also directly relates the stellar mass $M_*$
remaining in a dSph to its mean metallicity $\langle{\rm [Fe/H]}\rangle$
through the efficiency of outflows governed by the mass $M_h$ and the
radius $r_h$ of the dark matter halo. The observed relationship between
$\log(M_*/M_\odot)$ and $\langle{\rm [Fe/H]}\rangle$ has a slope of 2.5
over the wide range of $4.8\times 10^3\leq M_*/M_\odot\leq 4.6\times 10^8$
for dwarf galaxies (see SI Text).
This indicates $r_h\propto M_h^{1/3}$ in agreement with
the framework of hierarchical structure formation.
Our results confirm the previous studies of \cite{dekel} and \cite{woo}, and 
are in support of the early work of \cite{silk} on the general model of dwarf 
galaxy formation. 

Our model also demonstrates that for slow infall rates 
$\lambda_{\rm in}\ll\lambda$, the resulting MD 
must be extremely sharply peaked (essentially concentrated at a single 
value of [Fe/H]). Such MDs apply to all globular clusters with
the exception of the most massive ones (e.g., \cite{norris}).
This suggests that globular clusters are the result of 
the general process of astration governed by very slow infall rates 
compared to net gas loss rates. These slow infall rates further suggest that
globular clusters might be weak feeders during the inhomogeneous
evolution inside a large system and that they are not responsible for 
significant depletion of the baryonic supply of that system. As such,
the dark matter halo would be associated with the entire system, but
not be specifically tied to the globular clusters.
This is in contrast to dSphs, which result from processing all
the baryonic supply in their dark matter halos and are naturally
associated with these halos. Nonetheless, from the point of view of the 
analysis presented here, we consider globular clusters to be part of 
a family of early-formed dwarf galaxies but with low effective infall 
rates compared to outflow rates. With regard to the diverse morphological
types of dwarf galaxies, we consider dSphs to 
represent isolated evolution without dynamic effects from mergers or
tidal interactions with nearby systems. Other morphological
types may result from such effects. Insofar as all dwarf galaxies are 
related by the same general process as presented here, the observed 
ongoing astration in some dwarf irregular galaxies (e.g., \cite{evan})
poses a problem. 
These systems must be experiencing secondary processes of gas 
accretion due to local infall or mergers. Their MDs should have
a second peak due to the late infall.

The analysis presented here assumes that there is not a discontinuous 
onset of SNe Ia contributing Fe. Otherwise, the general outcome would
be MDs with two peaks, which are at most only rarely observed. 
It is not appropriate to extend this analysis to other metals than Fe 
without a more realistic treatment of the relative contributions of CCSNe
and SNe Ia that each produce very different yields of the other metals.
Such a treatment will be carried out in a subsequent paper that takes
into account the detailed distribution of the delay between the birth 
and death of the progenitors for SNe Ia. We note here that
as the rate of CCSNe decreases with decreasing gas mass, 
the contributions to Fe from SNe Ia will become larger or dominant. 
This is because previously-formed stars of low to intermediate masses 
in binaries will continue to evolve and produce
SNe Ia even after the SFR decreases. 
Crudely speaking, the net Fe production rate appropriate for later times
must be changed from Eq.~\ref{eq-pfe} to 
$P_{\rm Fe}(t)\sim X_{\rm Fe}^\odot
[\lambda_{\rm Fe}^{\rm CC}M_g(t)+\lambda_{\rm Fe}^{\rm Ia}M_g(t-\Delta)]$,
where $\lambda_{\rm Fe}^{\rm CC}$ and $\lambda_{\rm Fe}^{\rm Ia}$ are
the rate constants for CCSNe and SNe Ia, respectively, and $\Delta$ is
a typical delay between the birth and death of the progenitors for SNe Ia.
As SNe Ia do not produce e.g., O, Mg, and Si,
this will result in lower values of [E/Fe] for these elements 
at higher values of [Fe/H], which is in 
agreement with the general trend observed in dSphs 
(e.g., \cite{kirby11b}).

In conclusion, dSphs have evolved as a result of very massive 
gas loss and this gas has gone into the medium outside the dark matter 
halos associated with these galaxies. This means that these dark matter
halos must have masses $\sim 10^2$--$10^3$ times greater than the present 
total mass in stars after the gas loss and the cosmic ratio of baryonic to all 
matter are taken into account (see Table~S1). This is in qualitative 
accord with the conclusions from previous studies to infer the halo masses 
from observations of dSphs (e.g., \cite{wolf}).
As a result of the massive gas loss, extensive enrichment of metals has 
occurred in the general hierarchical structures outside individual dSphs. 
This drastically alters the subsequent chemical evolution of the 
emerging larger galaxies as most baryonic matter must have passed 
through processing in dSphs. As more massive dark matter halos
are formed during hierarchical growth, the efficiency of outflows will decrease.
Nevertheless, the general IGM must have been enriched by the net outflows 
from all dSphs. If the average IGM has [Fe/H]~$\sim -3$ and the outflows
from dSphs have [Fe/H]~$\sim -1.5$ on average, this would imply 
that $\sim 3\%$ of all baryonic matter was processed in dSphs.

\begin{acknowledgments}
We greatly appreciate the thoughtful and helpful comments by D. Lynden-Bell 
and R. Blandford in their efforts to improve our paper. 
We thank Evan Kirby for providing his data on the MDs of dSphs in 
electronic form.
This work was supported in part by DOE grant DE-FG02-87ER40328
(YZQ). GJW acknowledges NASA's Cosmochemistry Program
for research support provided through J. Nuth at the Goddard Space
Flight Center. He also appreciates the generosity of the Epsilon Foundation.
\end{acknowledgments}

\end{article}

\begin{figure*}
\centerline{\includegraphics[scale=0.7]{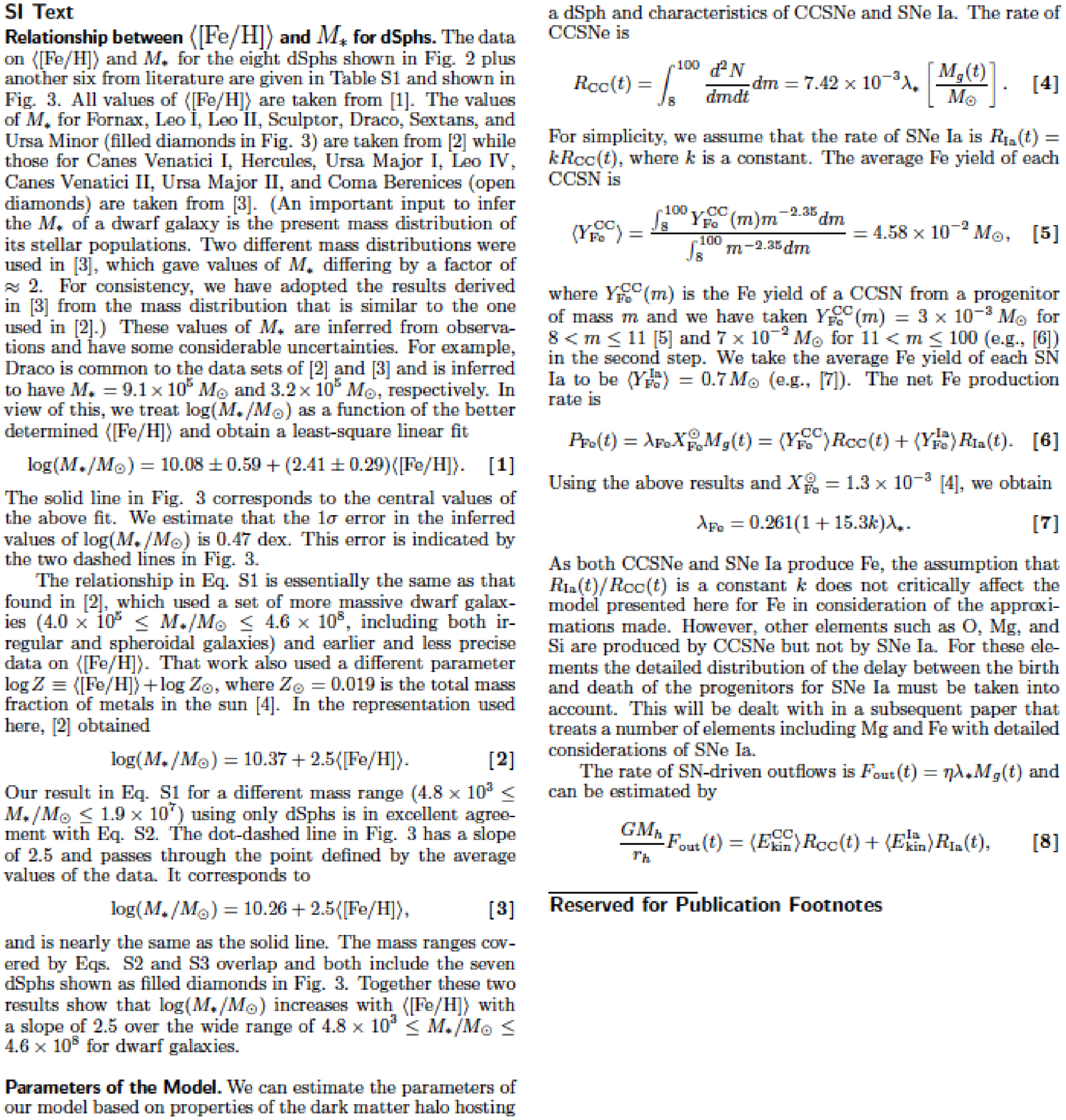}}
\end{figure*}

\begin{figure*}
\centerline{\includegraphics[scale=0.7]{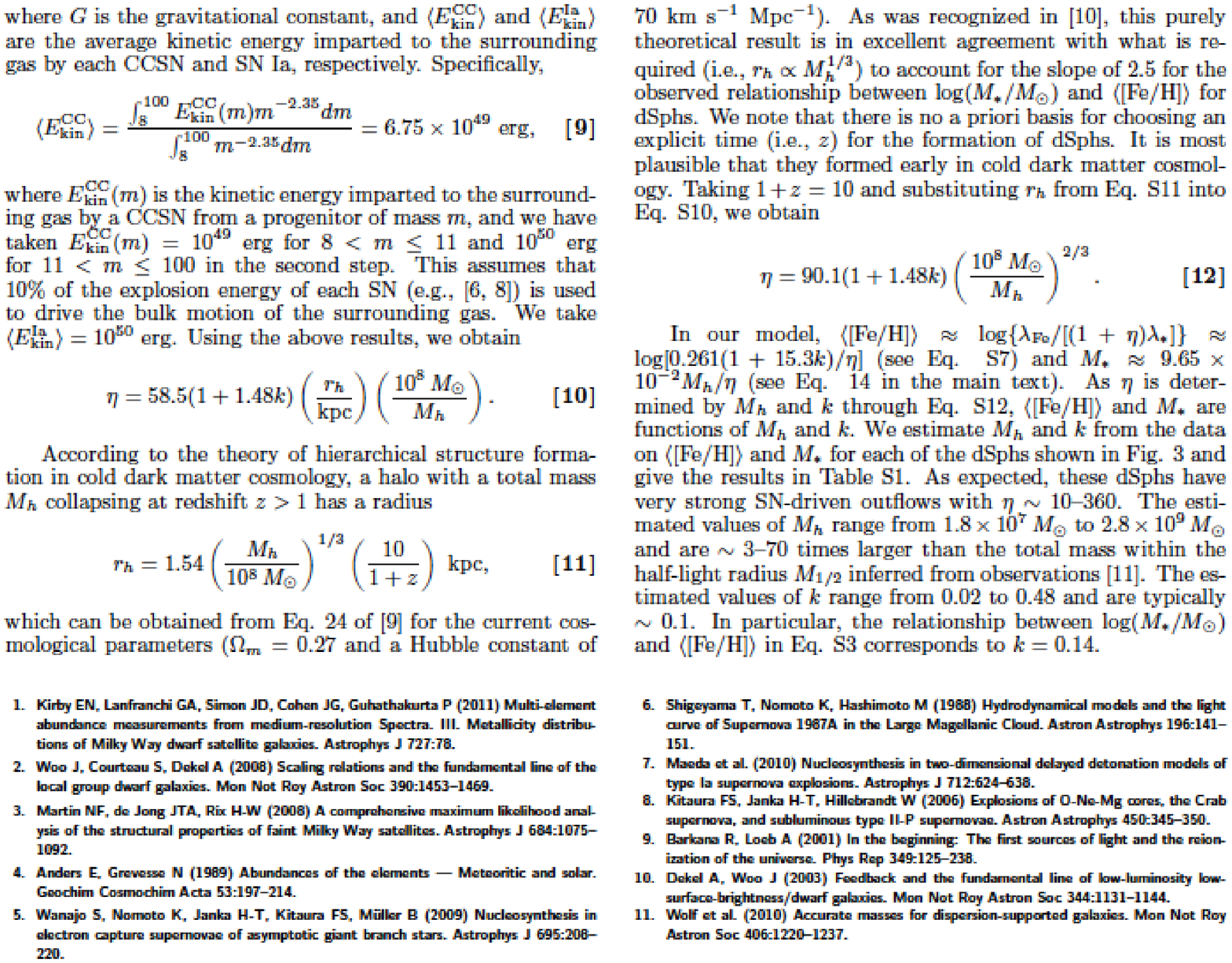}}
\end{figure*}

\begin{figure*}
\centerline{\includegraphics[scale=0.7]{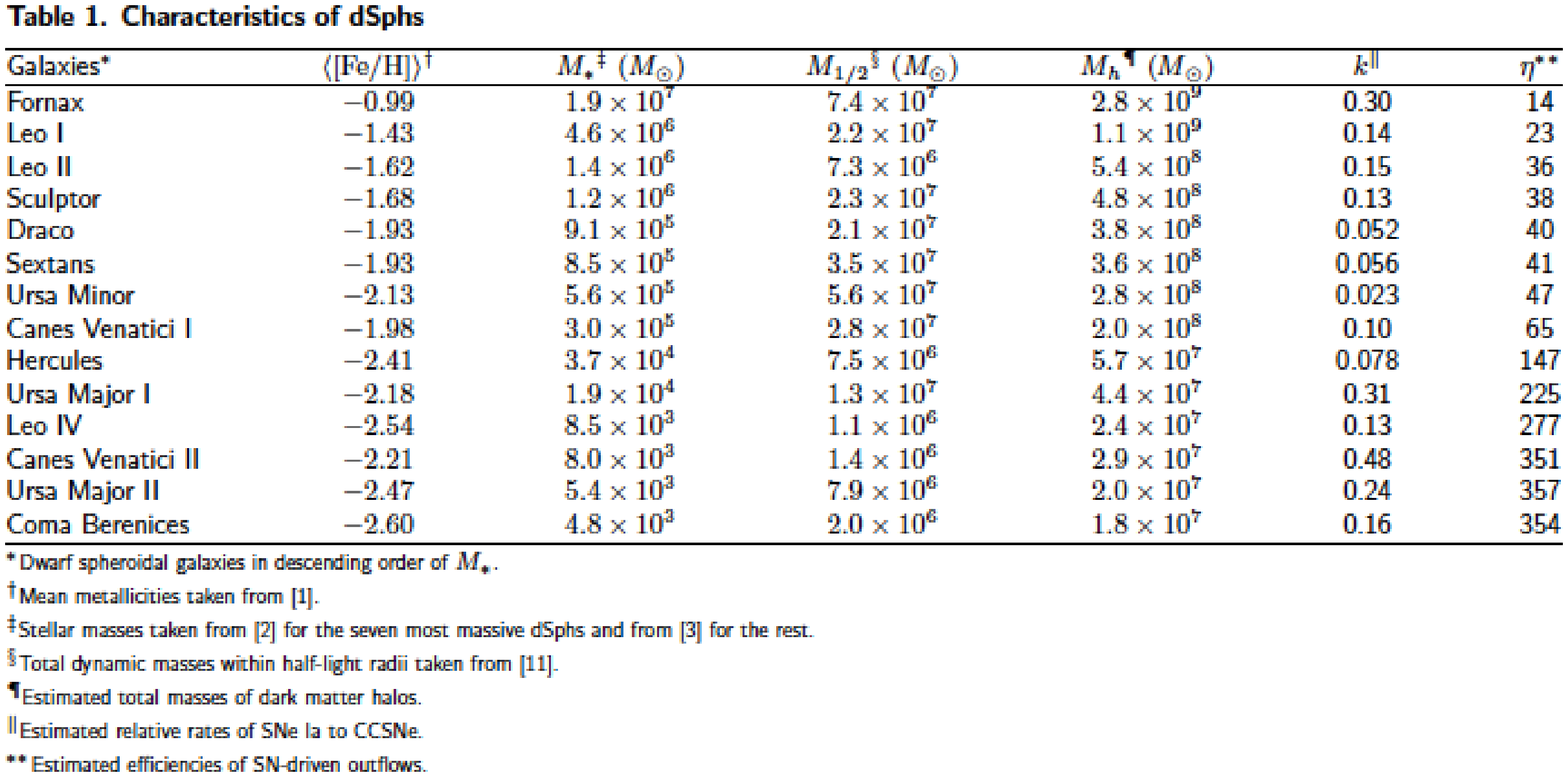}}
\end{figure*}

\end{document}